\documentclass[reqno,12pt]{amsart} 

\usepackage{amsmath}
\usepackage{amsfonts}
\usepackage{enumerate}
\usepackage{fancybox}

\theoremstyle{plain}
\newtheorem{thm}{Theorem}[section]{\bf}{\it}
\newtheorem{prop}[thm]{Proposition}{\bf}{\it}
\newtheorem{lemma}[thm]{Lemma}{\bf}{\it}
\newtheorem{cor}[thm]{Corollary}{\bf}{\it}
\theoremstyle{definition}
\newtheorem*{acknowledgement}{Acknowledgement} 
\newtheorem{remark}[thm]{Remark}{\it}{\rm}
\newtheorem{example}[thm]{Example}

\numberwithin{equation}{section}

\newcommand{\R}{{\mathbb R}}

\title[Positivity and decay of the electron density]{Positivity and
      lower bounds to the decay of the atomic one-electron density} 

\thanks{\copyright\ 2006 by the
        authors. This article may be reproduced in its entirety for
        non-commercial purposes.}

\author[S. Fournais, M. and T. Hoffmann-Ostenhof, and T. \O. S\o rensen]
       {S. Fournais \and M. Hoffmann-Ostenhof \and T. Hoffmann-Ostenhof \and
        T. \O stergaard S\o rensen}

\address[S. Fournais]{CNRS and Laboratoire de Math\'{e}matiques,
         Universit\'{e} Paris-Sud - B\^{a}t 425,
         F-91405 Orsay Cedex, France.}
\email{soeren.fournais@math.u-psud.fr}

\address[T. \O stergaard S\o rensen]{Laboratoire de Math\'{e}matiques,
         Universit\'{e} Paris-Sud - B\^{a}t 425,
         F-91405 Orsay Cedex, France.}

\address[T. \O stergaard S\o rensen, permanent address]
        {Department of Mathematical Sciences,
         Aalborg University,
         Fredrik Bajers Vej 7G,
         DK-9220 Aalborg East, Denmark.}
\email{sorensen@math.aau.dk}

\address[M. Hoffmann-Ostenhof]
        {Fakult\"at f\"ur Mathematik,
         Universit\"at Wien,         
         Nordbergstra\ss e 15,
         A-1090 Vienna, Austria.} 
\email{maria.hoffmann-ostenhof@univie.ac.at}

\address[T. Hoffmann-Ostenhof]{Institut f\"ur Theoretische
         Chemie, W\"ahringer\-strasse 17,
         Universit\"at Wien,
         A-1090 Vienna, Austria.}

\address[T. Hoffmann-Ostenhof, 2nd address]{
         The Erwin Schr\"{o}dinger International Institute for 
         Mathematical Physics,
         Boltzmanngasse 9, A-1090 Vienna, Austria.}
\email{thoffman@esi.ac.at}

\date{\today}

\begin{document}

\thispagestyle{empty}

\begin{abstract}
We investigate properties of the spherically averaged atomic one-electron 
density $\widetilde \rho(r)$. For a $\widetilde \rho$ which stems from
a physical ground state we prove that $\widetilde \rho>0$. We also give
exponentially decreasing lower bounds to $\widetilde \rho$ in the case
when the eigenvalue is below the corresponding essential spectrum.
\end{abstract} 

\maketitle

\section{Introduction and results}
Let $H$ be the non-relativistic Schr\"odinger operator of an
$N$-electron atom with nuclear charge $Z$ in the fixed nucleus  
approximation,
\begin{equation}\label{H}
  H=\sum_{j=1}^N\Big(-\Delta_j-\frac{Z}{|x_j|}\Big)
    +\sum_{1\le i<j\le N}\frac{1}{|x_i-x_j|}. 
\end{equation}
Here the $x_j=(x_{j,1},x_{j,2},x_{j,3})\in \mathbb R^3$, $j=1,\dots,
N$, denote the positions of the electrons, and the $\Delta_j$ are the
associated Laplacians so that $\Delta=\sum_{j=1}^N\Delta_j$ is the
$3N$-dimensional Laplacian. Let ${\bf x}=(x_1,x_2,\dots, x_N)\in
\mathbb R^{3N}$ and let $\nabla=(\nabla_1,\dots, \nabla_N)$ 
denote the $3N$-dimensional gradient operator. We write
\begin{align}\label{eq:HV}  
  H=-\Delta+V
\end{align}
 where $V$ is the multiplicative potential 
\begin{equation}\label{Vpot}
  V({\bf x})=\sum_{j=1}^N -\frac{Z}{|x_j|}+\sum_{1\le i<j\le
    N}\frac{1}{|x_i-x_j|}. 
\end{equation}
The operator $H$ is selfadjoint with operator domain $\mathcal
D(H)=W^{2,2}(\mathbb R^{3N})$ and form domain $\mathcal
Q(H)=W^{1,2}(\mathbb R^{3N})$ \cite{katobook}.

For an eigenfunction $\psi\in L^2(\R^{3N})$ of $H$, with eigenvalue
$E$, i.e., 
\begin{align}\label{eigen}
  H\psi=E\psi,
\end{align} 
we associate the {\it one-electron density} $\rho\in L^1(\R^3)$. It 
is defined by
\begin{equation}\label{rhohat}
  \rho(x)=\sum_{j=1}^N \rho_j(x)= \sum_{j=1}^N \int_{\mathbb
    R^{3N-3}}|\psi(x,\mathbf{\hat{x}}_j)|^2\,d\mathbf{\hat{x}}_j\,, 
\end{equation}
where we use the notation 
\begin{align*}
  {\mathbf{\hat{x}}_j}:=(x_1,\dots,x_{j-1},x_{j+1},\dots,
  x_N)\in\R^{3N-3}, 
\end{align*}
and
\begin{align*}
  d{\mathbf{\hat{x}}_j}:=dx_1\dots dx_{j-1}dx_{j+1}\dots dx_N,
\end{align*}
and, by abuse of notation, we identify
$(x_1,\dots,x_{j-1},x,x_{j+1},\dots,x_N)$ with
$(x,\mathbf{\hat{x}}_j)$. The spherically averaged density
\(\widetilde\rho\in L^1(\R_{+};r^2dr)\) is then defined by 
\begin{align}\label{def:rhoTilde}
   \widetilde{\rho}(r) &= \sum_{j=1}^N \widetilde{\rho}_j(r)=
   \sum_{j=1}^{N}\int_{{\mathbb S}^2}\rho_j(r\omega)\,d\omega\,, 
\end{align}
where \(r=|x|, \; \omega=x/|x|\in\mathbb S^2\) for \(x=r\omega\in\R^3\).

We assume throughout when studying \(\rho\) that \(E\) and \(\psi\) in
\eqref{eigen} are such that there exist constants \(C_0,\gamma>0\)
such that  
\begin{align}\label{eq:exp-dec}
  |\psi({\bf x})|\leq C_0\, e^{-\gamma|{\bf x}|}\quad\text{ for all }
  {\bf x}\in\mathbb R^{3N}.
\end{align}
The {\it a priori} estimate \cite[Theorem 1.2]{AHP} (see also
\cite[Remark 1.7]{AHP}) and \eqref{eq:exp-dec} imply the existence of
constants \(C_{1}, \gamma_{1}>0\) such that
\begin{align}\label{eq:dec_grad_psi}
  \big|\nabla\psi(\mathbf x)\big|\leq C_{1}\,e^{-\gamma_{1}|\mathbf
    x|} \ \text{ for almost all } \mathbf x\in\mathbb R^{3N}.
\end{align}
\begin{remark}\label{rem:exp-dec}
Since \(\psi\) is continuous (see \cite{Kato57}),
\eqref{eq:exp-dec} is only an assumption on the behaviour at infinity.
For references on the exponential decay of eigenfunctions, see e.g.\ 
\cite{FroeseHerbst} and \cite{Si-semi}.
The proofs of our results rely (if not indicated otherwise) on some
kind of decay-rate for \(\psi\); exponential decay is not essential,
but assumed for convenience. Note that \eqref{eq:exp-dec} and
\eqref{eq:dec_grad_psi} imply that \(\rho\) is Lipschitz continuous in
\(\R^3\) by Lebesgue's theorem on dominated convergence. This on the
other hand implies that \(\widetilde\rho\) is Lipschitz continuous in
\([0,\infty)\). 
\end{remark}
Since we are interested in atoms (in this non-relativistic
description with fixed nucleus) we have to take into account that
electrons are fermions. We shall work in the spin-independent
description. That is, we split $N$ such that 
$$
  N=N_1+N_2, \quad N_1,N_2\ge 0\,,
$$
and proceed as follows:
We associate Sobolev-spaces to this splititng.
Let \(\mathcal{S}(\R^{3N})\) be the space of Schwartz-functions, and
define 
\begin{align*}
  &\mathcal{S}_{N_1,N_2}(\R^{3N})=\{\varphi\in
  \mathcal{S}(\R^{3N})\:|\;\varphi(x_1,x_2,x_2, 
  \dots, x_{N_1},x_{N_1+1},\dots, x_N)\\\nonumber
  &\qquad\quad\text{is antisymmetric with respect to the first 
  $N_1$  coordinates}\\
  &\qquad\qquad\quad \text{ and antisymmetric in the remaining $N_2$
  coordinates.\,}\}
  \nonumber
\end{align*}
Therefore, for instance, 
\begin{align*}
  \varphi(x_1,&\dots,x_i,\dots, x_j,\dots,x_{N_1},\dots,x_N)
  \\&=
  {}-\varphi(x_1,\dots,x_j,\dots,x_i,\dots,x_{N_1},\dots,x_N).
\end{align*}
Similarily $\varphi$ changes sign if we interchange the coordinates of 
two electrons which belong to the other group of $N_2$ electrons which
are labeled with $i=N_1+1,\dots, N$. Note that in physical terms this
requirement means 
that the total spin is $\pm|N_2-N_1|/2$.
Define finally the Sobolev-spaces \(W^{p,q}_{N_1,N_2}(\R^{3N})\) as
the closure in the \(W^{p,q}(\R^{3N})\)-norm of
\(\mathcal{S}_{N_1,N_2}(\R^{3N})\).  

Let $H_{N_1,N_2}$ be the atomic $N$-electron Schr\"odinger operator
defined by \eqref{H}, restricted to functions with the above
symmetry. Then \(H_{N_1,N_2}\) has operator domain
\(\mathcal{D}(H_{N_1,N_2})=W^{2,2}_{N_1,N_2}(\R^{3N})\) and form domain
\(\mathcal{Q}(H_{N_1,N_2})=W^{1,2}_{N_1,N_2}(\R^{3N})\). 
We denote $E_{N_1,N_2}$ the infimum of its spectrum (when this is an
eigenvalue), and call it {\it the ground state energy}.  A
corresponding eigenfunction \(\psi=\psi_{N_1,N_2}\) is called a {\it
  ground state}. \(E\) will henceforth denote any eigenvalue.

Here is the first of our main results.
\begin{thm}\label{rho>0}
Let $\psi$ be a ground state of $H_{N_1,N_2}$, i.e., $H_{N_1,N_2} \psi
= E_{N_1,N_2} \psi$, and let $\widetilde \rho$ be the associated
spherically averaged density defined by
\eqref{rhohat}--\eqref{def:rhoTilde}. 

Then 
\begin{align}\label{eq:positivity}
  \widetilde\rho(r)>0\ \text{ for all } r\in[0,\infty)\,.
\end{align}
\end{thm}
\begin{remark}
  \(\, \)
\begin{enumerate}
\item[\rm (i)] 
At the origin we derive an explicit, positive lower bound to the density
(see \eqref{lb})
\begin{align}
  \rho(0)\ge \frac{2P^4}{3\pi ZN\|\psi\|^2}\quad,\quad \text{ with }
  P=\Big\|\sum_{j=1}^{N}\nabla_{j}\psi\Big\|. 
\end{align}
\item[\rm (ii)] Note that the choice of {\it anti-symmetric} in both
  groups of coordinates in the definition of
  \(\mathcal{S}_{N_1,N_2}(\R^{3N})\) is, in fact, not essential. One
  could consider functions {\it symmetric} in each group of
  coordinates. In fact, the theorem holds for any combination of
  sym\-metric/anti-symmetric. This will be clear from the proof. In
  particular, with \(N_1=N\) and symmetric (\(N_2=0\)), one gets the
  known result for the absolute (bosonic) ground state. 
\item[\rm (iii)]
  We would expect that the {\it non-averaged} density  $\rho$ of a
  ground state  of $H_{N_1,N_2}$ does not vanish either. Also,  the
  one-elec\-tron density \(\rho\) associated to fermionic ground
  states of {\it molecules} should be strictly positive.  However,
  these are much harder problems and they will not be addressed here. 
\item[\rm (iv)] It is not clear what to expect for excited
  states. Consider for instance a two-electron atom with no
  interelectronic repulsion; that is,
  $H=-\Delta_1-\Delta_2-Z/|x_1|-Z/|x_2|$. Let $\phi_i$, $i=1,2$, be
  linearly independent eigenfunctions of the three-dimensional
  one-electron operator $-\Delta-Z/|x|$ satisfying
  $\phi_1(0)=\phi_2(0)=0$. Then
  $\psi(x_1,x_2)=\phi_1(x_1)\phi_2(x_2)-\phi_1(x_2)\phi_2(x_1)$ is an
  eigenfunction of $H_{2,0}$ such that the associated density
  satisfies $\rho(0)=0$. However, it is not clear whether or not
  \(\rho\) still vanishes once the interelectronic repulsion is turned
  on. 
\end{enumerate}
\end{remark}
Our next result on the density is in the spirit of \cite{froese}. 
\begin{thm}\label{thm:lowerExpDensitySym}
Let $\psi$ be an eigenfunction of $H_{N_1,N_2}$  with eigenvalue $E$
and let \(\widetilde\rho\) be the associated spherically averaged
density defined by \eqref{rhohat}--\eqref{def:rhoTilde}. Define  
\begin{align}\label{eq:DefAlpha}
  \alpha_0 = \sup\big\{\alpha\,\big|\, e^{\alpha|{\bf x}|} \psi \in
  L^2({\mathbb R}^{3N})\big\}.
\end{align}

Then
\begin{align}\label{eq:limsupSym}
  \limsup_{R\rightarrow +\infty} 
  \Big(\frac{\ln\widetilde{\rho}(R)}{R}\Big)
  \leq -2\alpha_0.
\end{align}

If furthermore $E < \inf \sigma_{{\rm ess}}(H_{N_1,N_2})$, then also  
\begin{align}\label{eq:liminfSym}
  \liminf_{R\rightarrow +\infty} 
  \Big(\frac{\ln\widetilde{\rho}(R)}{R}\Big)
  \geq -2\sqrt{N} \alpha_0.
\end{align}
\end{thm}
\begin{remark} 
One can make these bounds more explicit using \cite{froese}; in fact,  
\begin{align}
  \alpha_0^2\le |E|\,.
\end{align}
To see this, we use Theorems 1.1 and 1.2 in \cite{froese}. The set of
thresholds $\mathcal T(H)$ is defined as the closure of the set of
eigenvalues of subsystems, i.e., the corresponding ionized systems. We
have, according to \cite{froese},  
\begin{equation} 
  \alpha_0^2+E\in \mathcal T(H)
  \ \text{ and }\ 
  \mathcal T(H)\subset (-\infty,0]\,,
\end{equation}
so that 
\begin{equation}\label{thresh}
  \alpha_0^2\le \sup\mathcal T(H)-E=|E|.
\end{equation}
\end{remark}

We shall discuss in Section~\ref{sec:disc} why the difference beween
the upper and lower bounds \eqref{eq:limsupSym}, \eqref{eq:liminfSym}
is to be expected. 

\begin{remark}\label{rem:genNbody}
It will be clear from the proof of
Theorem~\ref{thm:lowerExpDensitySym} (see Section~\ref{proof_exp_dec})
that the result  holds for more general $N$-body operators. That is,
we could  replace $V$ in \eqref{Vpot} by any multiplication operator
of the form $\sum_{\nu} v_{\nu}(\Pi_{\nu} {\bf x})$, ${\bf
  x}\in{\mathbb R}^{3N}$, where $\Pi_{\nu}$ is the orthogonal
projection of ${\mathbb R}^{3N}$ on a $d_{\nu}$-dimensional subspace,
and the $v_{\nu}$ are real-valued functions on ${\mathbb R}^{d_{\nu}}$
satisfying that \(v_{\nu} (-\Delta_{d_{\nu}}+1)^{-1}\) and
\((-\Delta_{d_{\nu}}+1)^{-1} (y\cdot \nabla
v_{\nu}(y))(-\Delta_{d_{\nu}}+1)^{-1}\) are compact as operators on
\(L^2({\mathbb R}^{d_{\nu}})\). Here $-\Delta_{d_{\nu}}$ is the usual
Laplace operator on ${\mathbb R}^{d_{\nu}}$, and \(y\in{\mathbb
  R}^{d_{\nu}}\) (compare with (1.2), (1.3) and (1.4) in \cite{froese}).
\end{remark}
The above theorem gives upper and lower exponential bounds on
$\widetilde{\rho}$ near infinity. Combined with Theorem~\ref{rho>0}
this implies (by continuity of $\widetilde{\rho}$, see
Remark~\ref{rem:exp-dec}) {\it global} lower exponential bounds in the
case of a ground state. We state this explicitely in the next
corollary.  
\begin{cor}
Let $\psi$ be an eigenfunction of $H_{N_1,N_2}$  with eigenvalue $E$, let
\(\alpha_0\) be as in \eqref{eq:DefAlpha}, and let \(\widetilde\rho\)
be the associated spherically averaged density defined by
\eqref{rhohat}--\eqref{def:rhoTilde}. 

If $E < \inf \sigma_{{\rm ess}}(H_{N_1,N_2})$, then for all
\(\alpha>\alpha_0\) there exists \(r_0\ge0\) and \(c=c(\alpha,r_0)>0\)
such that 
\begin{align}\label{eq:globalExp}
  \widetilde\rho(r)\geq c\,e^{-2\sqrt{N}\alpha r} \text{ for all }
  r\geq r_0\,.
\end{align}

If furthermore \(E=E_{N_1,N_2}\) (the ground state energy), then
\eqref{eq:globalExp} holds with $r_0=0$. 
\end{cor}

More detailed results than \eqref{eq:limsupSym} on upper bounds to
$\widetilde \rho$ are known. For completeness, we include the
following classical result; see \cite{AHHM:1981}.   
\begin{prop}\label{ubound}
Let $\psi$ be an eigenfunction of $H_{N_1,N_2}$  with eigenvalue $E$,
satisfying  
\begin{align}
  \epsilon:=\inf\sigma_{\rm ess}(H_{N_1,N_2})-E>0\,,
\end{align}
and let \(\rho\) be the associated density defined by
\eqref{rhohat}.

Then for all $r_0>0$ there exists a constant $C=C(r_0)>0$ such that
\begin{equation}\label{rhoup}
  \rho(x)\le C\,|x|^{\frac{Z-(N-1)}{\sqrt\epsilon}}e^{-2\sqrt\epsilon
  |x|}\ \text{ for all } x\in \mathbb R^3\text{ with }|x|\ge r_0\,.
\end{equation}
\end{prop}
\begin{remark}
  \(\, \)
\begin{enumerate}
\item[\rm (i)]
  The infimum of the essential spectrum of $H_{N_1,N_2}$ is
  characterized by the HVZ-theorem \cite[Theorem XII.17]{RS4}, which
  takes symmetry into account, in particular the fact that we consider
  fermions. Hence,
  $$
    \epsilon \ge   \min\big\{E_{N_1-1,N_2}, E_{N_1, N_2-1}\big\}-E.
  $$ 
\item[\rm (ii)]
  Note that the bound \eqref{rhoup} can be asymptotically sharp as has
  been shown for the ground state density of the Helium-like atom
  \cite{AHHM:1981}, where the physical ground state eigenfunction can
  be chosen positive. 
\item[\rm (iii)]
  If we consider a bosonic ground state, \(H\psi=E_0\psi\), for an
  atomic Hamiltonian which happens to have \(E_0= \inf \sigma_{{\rm
  ess}}(H)\) then the associated density can decay like $e^{-\beta
  \sqrt r}$ for some suitable $\beta$ up to some sub-exponential
  factors; see \cite{HHS:1983}.  
\end{enumerate}
\end{remark}
\section{Proof of positivity}
\label{proof_positivity}
\begin{proof}[Proof of Theorem~\ref{rho>0}]  
Assume first for contradiction that 
\begin{equation}\label{assumption}
  \widetilde\rho(r_0)=0 \text{ for some }r_0>0.
\end{equation}
This implies  that \(\widetilde\rho_j(r_0)=0\) for all
\(j=1,\ldots,N\), and therefore, still for all \(j=1,\ldots,N\), that  
\begin{align*}
  |\psi({\mathbf x})|^2=0 \ \text{ if }\ {\mathbf x}\in
  N_j(r_0)=\big\{{\mathbf x}\in\mathbb
  R^{3N}\,\big|\,|x_j|=r_0\big\}. 
\end{align*}
Here we used \eqref{rhohat}, \eqref{def:rhoTilde}, and the continuity
of \(\psi\). This means that  
\begin{align*}
  \psi({\mathbf x})=0\ \text{ for }\ {\mathbf x}\in N(r_0),
\end{align*}
where
\begin{align*}
  N(r_0)=\bigcup_{j=1}^N N_j(r_0)
  =\bigcup_{j=1}^N\big\{{\mathbf x}\in\mathbb
  R^{3N}\,\big|\,|x_j|=r_0\big\}. 
\end{align*}
Let
\begin{align*}
  \mathbf \Omega_0\equiv\mathbf \Omega_0(r_0)=\big\{{\mathbf x}
  \in\mathbb R^{3N}\,\big|\, \max_{j=1,\ldots,N}|x_j|<r_0\big\}. 
\end{align*}
Then \(\mathbf \Omega_0\) is an open bounded subset of \(\mathbb
R^{3N}\) satisfying 
\begin{align*}
  {\mathbf x}\in\mathbf \Omega_0  \ \Rightarrow \ {\mathcal P}{\mathbf
    x}\in\mathbf \Omega_0
\end{align*}
for every permutation $\mathcal P\in\mathfrak S^N$ of the electron
coordinates (that is, of the \(N\)-tuple \((1,\ldots,N)\)). This means
that the following space is non-trivial (\(\neq\{0\}\)): 
\begin{align*}
  \mathcal Q_{N_1,N_2}(\mathbf \Omega_0)=
  \{ f \in W_0^{1,2}(\mathbf \Omega_0) \,\big| \, \exists F \in
  \mathcal Q(H_{N_1,N_2}) \text{ such that } F|_{\mathbf \Omega_0} = f
  \}. 
\end{align*}
In fact, by the above, \(\psi=0\) on \(\partial\mathbf \Omega_0\). 
Therefore, the restriction \(\psi_{\mathbf \Omega_0}:=\psi|_{\mathbf
  \Omega_0}\) of \(\psi\in\mathcal Q(H_{N_1,N_2})\) to \(\mathbf
\Omega_0\) belongs to \(\mathcal Q_{N_1,N_2}(\mathbf \Omega_0)\), and
clearly 
\begin{align}\label{eq:expect}
  \langle\psi_{\mathbf \Omega_0}, H\psi_{\mathbf \Omega_0}\rangle 
  = E_{N_1,N_2}\|\psi_{\mathbf \Omega_0}\|^2\,.
\end{align}

We now claim that we have the strict inequality
\begin{equation}\label{EOm0}
  E_{N_1,N_2}(\mathbf \Omega_0)=\min_{\varphi\in\mathcal
    Q_{N_1,N_2}(\mathbf \Omega_0)} 
  \frac{\langle \varphi,
  H\varphi\rangle}{\|\varphi\|^2}>E_{N_1,N_2}=\min_{\varphi\in\mathcal 
  Q(H_{N_1,N_2})}\frac{\langle \varphi, H\varphi\rangle}{\|\varphi\|^2}. 
\end{equation}
Indeed, assume that we have equality in \eqref{EOm0}. Then, by the
variational characterization of the ground state (also in a symmetry
subspace), the eigenfunction which minimizes
the LHS of \eqref{EOm0}, and which we extend to \(\mathbb R^{3N}\) by
setting it identically equal to zero outside $\mathbf \Omega_0$, will
have to be  an eigenfunction in all of $\mathbb R^{3N}$ also. This is
a contradiction to unique continuation (see \cite[Theorem XIII.57]{RS4}).

Now, \eqref{eq:expect} and \eqref{EOm0} imply that \(\psi|_{\mathbf
  \Omega_0}=0\). By unique continuation, \(\psi=0\) (since
\(r_0>0\)). This is a contradiction, and therefore settles the case
when \(r_0>0\) (see \eqref{assumption}). 

We still have to show that 
\begin{equation}\label{l0bound}
  \widetilde\rho(0)=4\pi\rho(0)>0.
\end{equation} 
Here we explicitely use the Coulombic nature of the potential \(V\)
(see \eqref{Vpot}). Recall that \(\mathbf x=(x_1,\ldots,x_N)\in\mathbb
R^{3N}\), with \(x_{j}=(x_{j,1}, x_{j,2},x_{j,3})\in\mathbb
R^{3}\). The gradient with respect to \(x_j\) is denoted \(\nabla_j\).  

Let, for \(\alpha\in\mathbb R\), 
\begin{equation}\label{Falpha}
  F^{(\alpha)}=\sum_{j=1}^N(\nabla_j+\alpha x_j)\psi
  =\big(F^{(\alpha)}_1,F^{(\alpha)}_2,F^{(\alpha)}_3\big)
\end{equation}
with \(F^{(\alpha)}_k=\sum_{j=1}^N\big(\frac{\partial\psi}{\partial
  x_{j,k}}+\alpha x_{j,k}\psi\big), k=1,2,3\). Using that
\(\psi\in{\mathcal D}(H)=W^{2,2}(\R^{3N})\), and the exponential decay
\eqref{eq:exp-dec} and \eqref{eq:dec_grad_psi}, we get that
\(F^{(\alpha)}_k\in W^{1,2}(\mathbb R^{3N})\). Then the following
variational expression is well-defined: 
\begin{equation}\label{varF}
  \mathbf R(\alpha)=\sum_{k=1}^{3}\big\langle F^{(\alpha)}_k,
  (H-E_{N_1,N_2})F^{(\alpha)}_k\big\rangle.
\end{equation}
Note that  the \(F^{(\alpha)}_k\) are obtained by applying the
operator  
\begin{align*}
  \sum_{j=1}^{N}(\frac{\partial}{\partial x_{j,k}}+\alpha x_{j,k})
\end{align*}
to $\psi$, which does not change symmetry properties with respect to
permutations of the electron coordinates.
Hence $F^{(\alpha)}_k$ has the same permutational  properties as
$\psi$ itself, and so \(F^{(\alpha)}_k\in\mathcal Q(H_{N_1,N_2}),
k=1,2,3\). Therefore, by the variational characterization of the
ground state \(\psi\) (also in a symmetry subspace), we have
\begin{equation}\label{R>}
  \mathbf R(\alpha)\geq 0\ \text{ for all} \ \alpha\in\mathbb R.
\end{equation}
We have (since \(H\) is self-adjoint) that
\begin{align}\label{eq:R_alpha}
  {\mathbf R}(\alpha)&=\sum_{k=1}^{3}\sum_{i,j=1}^{N}\big(a_{i,j,k}
  +2\alpha b_{i,j,k} +\alpha^2c_{i,j,k}\big),\nonumber 
  \displaybreak[0]  
  \\
  a_{i,j,k}&=\big\langle\frac{\partial\psi}{\partial
  x_{j,k}},(H-E_{N_1,N_2})\frac{\partial\psi}{\partial 
  x_{i,k}}\big\rangle,\nonumber
  \displaybreak[0]  
  \\
  b_{i,j,k}&=\big\langle\frac{\partial\psi}{\partial
  x_{j,k}},(H-E_{N_1,N_2})(x_{i,k}\psi)\big\rangle,\nonumber
  \displaybreak[0]  
  \\
  c_{i,j,k}&=\big\langle(x_{j,k}\psi),(H-E_{N_1,N_2})(x_{i,k}\psi)\big\rangle.  
\end{align}
First note that, since \(H\psi=E_{N_1,N_2}\psi\), (\([\ ; \ ]\)
denoting the commutator) 
\begin{align*}
  \big(H-E_{N_1,N_2})\frac{\partial\psi}{\partial
  x_{i,k}}= \Big[(H-E_{N_1,N_2});\frac{\partial}{\partial
  x_{i,k}}\Big]\psi
  =\big(-\frac{\partial V}{\partial x_{i,k}}\big)\psi,
\end{align*}
and so, by partial integration, 
\begin{align*}
  \sum_{k=1}^{3}\sum_{i,j=1}^{N}a_{i,j,k}
  &=\sum_{k=1}^{3}\sum_{i,j=1}^{N}\big\langle\frac{\partial\psi}{\partial  
  x_{j,k}}, \big(-\frac{\partial V}{\partial
  x_{i,k}}\big)\psi\big\rangle
  \\&=\frac12\sum_{k=1}^{3}\sum_{i,j=1}^{N}\big\langle\psi,\frac{\partial^2V}{\partial
  x_{j,k}\partial x_{i,k}}\psi\big\rangle\,.
\end{align*}
Strictly speaking, \(V\) and \(\psi\) are not smooth enough for this
and the following computation to be but formal. However, regularizing
\(V\) and using the exponential decay
\eqref{eq:exp-dec}--\eqref{eq:dec_grad_psi}, and the continuity of
\(\rho\) (see Remark~\ref{rem:exp-dec}), one easily justifies this.

Note that (for \(V\), see \eqref{Vpot})
\begin{align*}
  &\big(\frac{\partial}{\partial
  x_{i,k}}+\frac{\partial}{\partial
  x_{j,k}}\big)\big(|x_i-x_j|^{-1}\big)=0,\\  
  &\sum_{k=1}^{3}\frac{\partial^2(|x_j|^{-1})}{\partial x_{j,k}\partial
  x_{i,k}} =
  \delta_{i,j}\Delta_{j}(|x_j|^{-1})=-4\pi\delta_{i,j}\delta(|x_j|),
\end{align*}
where \(\delta_{i,j}\) is Kronecker's delta, and \(\delta\) is the
delta - function. This way, 
\begin{align}
  \label{eq:a}
  \sum_{k=1}^{3}\sum_{i,j=1}^{N}a_{i,j,k}&=\sum_{j=1}^{N}\big\langle\psi,2\pi 
  Z\,\delta(|x_{j}|)\psi\big\rangle
  =2\pi Z\int_{\mathbb R^{3N}}|\psi(\mathbf
  x)|^2\delta(|x_j|)\,d\mathbf x\nonumber\\
  &=2\pi Z\sum_{j=1}^{N}\rho_j(0)=2\pi Z\rho(0). 
\end{align}
Secondly, again since \(H\psi=E_{N_1,N_2}\psi\),
\begin{align}
\label{eq:star}
  \big(H-E_{N_1,N_2})(x_{i,k}\psi)= \big[(H-E_{N_1,N_2});x_{i,k}\big]\psi
  ={}-2\frac{\partial \psi}{\partial x_{i,k}}.
\end{align}
Therefore,
\begin{align}\label{eq:b}\nonumber
  \sum_{k=1}^{3}\sum_{i,j=1}^{N}b_{i,j,k}&=
  {}-2\sum_{k=1}^{3}\big\langle\sum_{j=1}^{N}\frac{\partial
  \psi}{\partial x_{j,k}},\sum_{i=1}^{N}\frac{\partial
  \psi}{\partial x_{i,k}} \big\rangle
  \\&=-2\Big\|\sum_{j=1}^{N}\nabla_{j}\psi\Big\|^{2}=:-2P^2.
\end{align}
Finally, using \eqref{eq:star} and partial integration,   
\begin{align}\label{eq:c}\nonumber
  \sum_{k=1}^{3}\sum_{i,j=1}^{N}c_{i,j,k}&=
  -2\sum_{k=1}^{3}\sum_{i,j=1}^{N}\big\langle
  x_{j,k}\psi,\frac{\partial\psi}{\partial x_{i,k}}\big\rangle
  \\&=-\sum_{k=1}^{3}\sum_{i,j=1}^{N}\int_{\mathbb
  R^{3N}}x_{j,k}\frac{\partial}{\partial
  x_{i,k}}\big(\psi^2)\,d\mathbf x\nonumber\\
  &=\sum_{k=1}^{3}\sum_{i,j=1}^{N}\delta_{i,j}\cdot \|\psi\|^{2}
  =3N\|\psi\|^2.
\end{align}
Combining \eqref{eq:a}, \eqref{eq:b}, and \eqref{eq:c} with
\eqref{eq:R_alpha}, we find that 
\begin{align*}
  \mathbf R(\alpha)=2\pi Z\rho(0)-4\alpha P^2+3\alpha^2N\|\psi\|^2\geq 0 
  \ \text{ for all } \ \alpha\in\mathbb R.
\end{align*}
Optimizing this expression in \(\alpha\) we obtain
\begin{align}\label{lb}
  \rho(0)\ge \frac{2P^4}{3\pi ZN\|\psi\|^2}\quad,\quad
  P=\Big\|\sum_{j=1}^{N}\nabla_{j}\psi\Big\|. 
\end{align}
Suppose that $P=0$. After Fourier transformation, this implies that
\begin{align}\label{eq:Tnul}
  \sum_{j=1}^N p_j\,\hat{\psi}({\mathbf p}) =0,
\end{align}
in $L^2({\mathbb R}^{3N})$. The equation \eqref{eq:Tnul} clearly
implies that $\hat{\psi}=0$ (since it has support on the set
$\{\,{\mathbf p} \in {\mathbb R}^{3N} \,|\, \sum_{j=1}^N p_j=0\,\}$
which has zero measure). Since $\psi$ is an eigenfunction,
\(\psi\neq0\), so this is a contradiction. We conclude that $P \neq 0$
and \eqref{lb} thus implies \eqref{l0bound}.

This settles the case \(r_0=0\), and therefore finishes the proof of  
Theorem~\ref{rho>0}.
\end{proof}
\section{Discussion and proof of decay}
\label{proof_exp_dec}
\subsection{Discussion and examples}\label{sec:disc}
Before proving Theorem~\ref{thm:lowerExpDensitySym}, we show with some
examples why the difference on the right hand sides of
\eqref{eq:limsupSym} and \eqref{eq:liminfSym} is natural. Define
$\widetilde{\alpha}_0$ as 
\begin{align}
  \widetilde{\alpha}_0 = \sup\big\{\alpha\,\big|\, e^{2\alpha r}
  \widetilde{\rho}(r) \in L^1({\mathbb R}_{+}, r^2dr) \big\}.
\end{align}
For the definition of \(\alpha_0\), see \eqref{eq:DefAlpha}.
\begin{example}[$\widetilde{\alpha}_0=\alpha_0$]\label{ex:first}
Consider first the function,
$$
  \psi_1(x_1,x_2) = e^{-\alpha_1|x_1|} e^{-\alpha_2|x_2|}\quad,\quad
  x_1,x_2\in\mathbb R^{3}. 
$$
Clearly, the associated density satisfies,
$$
  \widetilde{\rho}(r) = c_1 e^{-2\alpha_1 r} + c_2 e^{-2\alpha_2
  r}\quad,\quad c_1=\pi/\alpha_2^3\ ,\  c_2=\pi/\alpha_1^3.
$$
Thus, $\widetilde{\alpha}_0 = \min(\alpha_1, \alpha_2)$. It is not
hard to see that in this example also $\alpha_0 = \min(\alpha_1,
\alpha_2)$: Clearly, $\alpha_0 \geq \min(\alpha_1,
\alpha_2)$. Suppose, without loss of generality, that
$\alpha_1\leq\alpha_2$. If $\alpha>\alpha_1$, then (by continuity)
there exists a neighbourhood $\Gamma$ (in ${\mathbb S}^5$) of
$(1,0,0,0,0,0)$, and $\epsilon > 0$, such that, for all
$(\omega_1,\omega_2) \in \Gamma$,  
$$
  \alpha -\alpha_1 |\omega_1| - \alpha_2 |\omega_2| > \epsilon.
$$
The integral of $|e^{\alpha |{\bf x}|} \psi|^2$ over the cone
${\mathbb R}_{+} \Gamma\subset\mathbb R^6$ therefore clearly
diverges. Thus,   
$$
  \alpha_0 = \min(\alpha_1, \alpha_2).
$$
\end{example}
\begin{example}[$\widetilde{\alpha}_0\neq\alpha_0$]\label{ex:second}
Consider now the function
$$
  \psi_2(x_1,x_2) = e^{-\alpha_1|\frac{x_1+x_2}{\sqrt{2}}|}
  e^{-\alpha_2|\frac{x_1-x_2}{\sqrt{2}}|}\quad,\quad
  x_1,x_2\in\mathbb R^{3}. 
$$
It is easy to see that $\alpha_0 = \min(\alpha_1, \alpha_2)$ also in
this case (the definition of $\alpha_0$ is invariant under an
orthogonal change of coordinates). However, we will see that
$\widetilde{\alpha}_0 = \sqrt{2}\min(\alpha_1, \alpha_2)$. We write
out 
\begin{align*}
  \int_{{\mathbb R}^3} e^{2\alpha |x_1|} \rho_1(x_1) \,dx_1 &=
  \int_{{\mathbb R}^6} e^{2\alpha |x_1|} 
  e^{-2\alpha_1|\frac{x_1+x_2}{\sqrt{2}}|}
  e^{-2\alpha_2|\frac{x_1-x_2}{\sqrt{2}}|}\,dx_1dx_2 \\
  &=
  \int_{{\mathbb R}^6} e^{2\alpha |\frac{y_1+y_2}{\sqrt{2}}|} 
  e^{-2\alpha_1|y_1|}
  e^{-2\alpha_2|y_2|}\,dy_1dy_2.
\end{align*}
Since $|y_1+y_2|\leq |y_1| + |y_2|$, the above integral is clearly
convergent for all $\alpha$ satisfying
$\frac{\alpha}{\sqrt{2}}<\min(\alpha_1, \alpha_2)$. It is also easy to
see (as in the previous example) that the integral is not convergent
(on a suitable cone) if $\frac{\alpha}{\sqrt{2}}>\min(\alpha_1,
\alpha_2)$. 
\end{example}
\begin{remark}
Example~\ref{ex:first} would be the correct behaviour (modulo
polynomial forefactors) of eigenfunctions of $H$, if we omitted the
terms $|x_j-x_k|^{-1}$ in $V$. Since our proof works for general
$N$-body operators (see Remark~\ref{rem:genNbody}), it is therefore
clear that Example~\ref{ex:second} is equally relevant, since
Example~\ref{ex:second} is obtained by using a rotation of the
coordinates in Example~\ref{ex:first}.  
\end{remark}
The proof of Theorem~\ref{thm:lowerExpDensitySym} will rely on the
result \cite[Theorem 2.1]{froese} adapted for our purpose.
For $R_1<R_2$,
denote by $\Omega(R_1,R_2)$ the spherical shell (in ${\mathbb
R}^{3N}$)
\begin{align}
  \label{eq:Omega}
  \Omega(R_1,R_2) = \big\{{\mathbf x}\in\mathbb R^{3N} \,\big|\, R_1
  \leq |{\mathbf x}| \leq R_2\big\}.
\end{align}
\begin{thm}\label{thm:fh21}
Suppose $\psi$ is an eigenfunction of $H$ with eigenvalue $E$
and let $\alpha_0$ be defined by
\eqref{eq:DefAlpha}. 
Suppose
$\delta(R)$ is a positive function satisfying 
\begin{align}\label{eq:delta}
  \liminf_{R\rightarrow +\infty}\Big(\frac{\ln \delta(R)}{R}\Big) \geq 0.
\end{align}

Then
\begin{align}\label{eq:thm31}
  \lim_{R\rightarrow +\infty}\frac{1}{R}\ln\Big(
  \int_{\Omega(R, R+\delta(R))}
  \!\!\!\!\!\!\!\!\!\!\!\!\!\!\!
  |\psi({\mathbf x})|^2\,d{\mathbf x}\Big)
  = -2\alpha_0.
\end{align}
\end{thm}
The next result, 
which is a special case of \cite[Theorem 2.2]{froese},
will not be used in the sequel. It is given only in order for the
reader to be able to compare the result for the density, 
Theorem~\ref{thm:lowerExpDensitySym}, with the corresponding result
for the spherically averaged eigenfunction. 
\begin{thm}\label{thm:fh23}
Suppose in addition to the hypotheses of Theorem~\ref{thm:fh21} that
$E< \inf \sigma_{{\rm ess}}(H)$. 

Then
\begin{align*}
  \lim_{R\rightarrow +\infty}\frac{1}{R} \ln
  \Big(\int_{{\mathbb S}^{3N-1}} |\psi(R\Omega)|^2 \, d\Omega\Big) =
  -2\alpha_0.
\end{align*}
\end{thm}
Before starting the proof of Theorem~\ref{thm:lowerExpDensitySym}, we
state and prove a result similar to Theorem~\ref{thm:fh21} for the
density \(\widetilde\rho\). 
\begin{thm}\label{thm:21bis}
Suppose $\psi$ is an eigenfunction of $H$ with associated electron
density $\widetilde{\rho}$ and let $\alpha_0$ be defined by
\eqref{eq:DefAlpha}. 

Then
\begin{align}
  \label{eq:uppersolid}
   &\limsup_{R\rightarrow +\infty}\frac{1}{R}
  \ln
  \Big(\int_{R}^{\infty}\widetilde\rho(r)r^2\,dr\Big)
  \leq  -2\alpha_0, \\
  \label{eq:lowersolid}
  &\liminf_{R\rightarrow +\infty}\frac{1}{R}
  \ln\Big(
  \int_{R}^{\infty}\widetilde\rho(r)r^2\,dr\Big)
  \geq
  -2\sqrt{N}\alpha_0\,.
\end{align}
\end{thm}
\begin{proof}
The proof of \eqref{eq:uppersolid} is a simple calculation:
\begin{align*}
  \int_R^{\infty} \widetilde{\rho}(r) r^2 \,dr
  &\leq N
  \int_{\{{\mathbf x} \,:\, \max_j|x_j| \geq R\}}
  |\psi({\mathbf x})|^2\,d{\mathbf x} \\
  &\leq N
  \int_{\{{\mathbf x} \,:\, |{\mathbf x}| \geq R\}}
  |\psi({\mathbf x})|^2\,d{\mathbf x}\,.
\end{align*}
Therefore, for all $\alpha < \alpha_0$,
\begin{align*}
  \int_R^{\infty} \widetilde{\rho}(r) r^2 \,dr &
  \leq N
  e^{-2\alpha R} \| e^{\alpha |{\mathbf x}|} \psi
  \|_{L^2({\mathbb R}^{3N})}^2\,.
\end{align*}
This clearly implies \eqref{eq:uppersolid}.

In order to prove \eqref{eq:lowersolid}, we calculate 
\begin{align*}
  \int_R^{\infty} &\widetilde{\rho}(r) r^2\,dr \\
  &=
  \sum_{j=1}^N \int_R^{\infty}r^2\,dr
  \int_{{\mathbb S}^2} d\omega
  \int_{{\mathbb R}^{3N}} |\psi(x_1,\ldots,x_N)|^2
  \delta(r\omega-x_j)\,d{\mathbf x} \\
  &=\sum_{j=1}^N
  \int_{\{|x_j| \in [R, \infty),\, \hat{\bf x}_j \in {\mathbb
  R}^{3N-3}\}} |\psi(x_1,\ldots,x_N)|^2 \,d{\mathbf x} \\
  &\geq\int_{\{{\mathbf x} \,:\, \max_j|x_j| \geq R \}}
  |\psi(x_1,\ldots,x_N)|^2 \,d{\mathbf x}\,. 
\end{align*}
Using
\begin{align*}
  \min_j |x_j| \leq \frac{|{\mathbf x}|}{\sqrt{N}} \leq \max_j|x_j|, 
\end{align*}
we get
\begin{align*}
  \big\{{\mathbf x}\in\mathbb R^{3N} \,\big|\, \sqrt{N}R\leq 
  |{\mathbf x}|  \big\} \subset  
  \big\{{\mathbf x}\in\mathbb R^{3N}\,\big|\, &\max_j |x_j| \geq R
  \big\}, 
\end{align*}
and so, from the above, 
\begin{align*}
  \int_R^{\infty} \widetilde{\rho}(r) r^2\,dr &\geq
  \int_{\{{\mathbf x} \,:\, \sqrt{N}R\leq |{\mathbf x}| \}}
  |\psi(x_1,\ldots,x_N)|^2 \,d{\mathbf x}\,. 
\end{align*}
The lower bound \eqref{eq:lowersolid}, now follows from
Theorem~\ref{thm:fh21} by taking \(\delta(R)=+\infty\).
\end{proof}
\subsection{Proof of
  Theorem~\ref{thm:lowerExpDensitySym}}\label{sec:proof} 
We now start the proof of Theorem~\ref{thm:lowerExpDensitySym}. It
will be based on a number of lemmas. 

We start by defining  an operator $T_R$ which maps functions from
$H^{1/2+\epsilon}({\mathbb R}^{3N})$ continuously to
$L^2(\mathbb{S}^2\times{\mathbb R}^{3N-3})$. $T_R$ is the restriction
map 
\begin{align}\label{def:T_R}
  [T_R f](\omega, \hat{\bf x}_1) = f(R\omega, \hat{\bf x}_1)\,,\quad
  R>0\, ,  
  \omega\in\mathbb{S}^2\, , \hat{\mathbf x}_1\in\R^{3N-3}\,.
\end{align}
Lemma~\ref{lem:A} will allow us to get the upper bound
\eqref{eq:limsupSym}.
\begin{lemma}\label{lem:A}
Suppose $\psi$ is an eigenfunction of $H$ with eigenvalue $E$. Then
there exists $c>0$ such that, for all $R>1$ and all $j \in \{1,
\ldots, N\}$, 
\begin{align}\label{eq:IndivTerms}
  \widetilde{\rho}_j(R) \leq c R^{-2} \int_{R-1}^{\infty}
  \widetilde{\rho}_j(r) r^2 \,dr\,.
\end{align}
In particular,
\begin{align}\label{eq:sumofterms}
  \widetilde{\rho}(R) \leq c R^{-2} \int_{R-1}^{\infty}
  \widetilde{\rho}(r) r^2 \,dr.
\end{align}
\end{lemma}
\begin{proof}
Since \eqref{eq:sumofterms} follows from \eqref{eq:IndivTerms} by
summation over $j$, it clearly suffices to prove the latter. Without
loss of generality, we will restrict attention to the case $j=1$. It
is well-known that the restriction map $T_R$ from \eqref{def:T_R}
defines a bounded map between Sobolev spaces with loss of a half
($+\epsilon$) derivative. We need some control of how the constants
depend on $R$, but not the optimal regularity result, so we state and
prove the following (elementary, not optimal) lemma. 
\begin{lemma}\label{lem:SobTrace}
The map $T_R$ defines a bounded operator from
$H^1({\mathbb R}^{3N})$ to $L^2(\mathbb{S}^2\times{\mathbb
  R}^{3N-3})$, with the estimate 
\begin{align}\label{eq:SobIneq}
  \| T_R f \|_{L^2(\mathbb{S}^2\times{\mathbb R}^{3N-3})}
  \leq R^{-1}\big( \| \nabla f \|_{L^2({\mathbb R}^{3N})} + \| f
  \|_{L^2({\mathbb R}^{3N})} \big)\,.
\end{align}
\end{lemma}
\begin{proof}[Proof of Lemma~\ref{lem:SobTrace}]
(The proof is a repetition of the proof of 
\newline\noindent
\cite[Lemma~3.1]{froese}).
For $f \in C_0^{\infty}({\mathbb R}^{3N})$, we have 
\begin{align*}
  &\int_{\mathbb{S}^2\times{\mathbb R}^{3N-3}}
  |f(R\omega,\hat{\bf x}_1)|^2\,d\omega\,d\hat{\bf x}_1 \\
  &=
  \int_R^{\infty}\int_{\mathbb{S}^2\times{\mathbb R}^{3N-3}}
  -\frac{d}{dr}
  |f(r\omega,\hat{\bf x}_1)|^2\,d\omega\,d\hat{\bf x}_1\,dr \\
  &=
  \int_R^{\infty}\int_{\mathbb{S}^2\times{\mathbb R}^{3N-3}}
  -2 \text{Re}\big\{\overline{f(r\omega, \hat{\bf x}_1)}\frac{d}{dr}
  f(r\omega,\hat{\bf x}_1)\big\} \,d\omega\,d\hat{\bf x}_1\,dr \\
  &\leq
  R^{-2}
  \int_R^{\infty}\int_{\mathbb{S}^2\times{\mathbb R}^{3N-3}}
  \big(|f(r\omega, \hat{\bf x}_1)|^2 +
  |\nabla f(r\omega, \hat{\bf x}_1)|^2 \big) r^2 \,d\omega
  \,d\hat{\bf x}_1\,dr\,.
\end{align*}
This clearly implies \eqref{eq:SobIneq}, from which
Lemma~\ref{lem:SobTrace} follows. 
\end{proof}
We now finish the proof of Lemma~\ref{lem:A}. Let $\chi\in
C^{\infty}({\mathbb R})$ be monotone, \(0\leq\chi\leq1\), such that
\begin{align}
  \chi(t) =0 \text{ for } t\leq 1/2\quad,\quad
  \chi(t) =1 \text{ for } t\geq 1.
\end{align}
Define $\chi_R({\mathbf x}) = \chi(|x_1|-(R-1))$. 
With $\psi$ being the eigenfunction of $H$, we have
$T_R\psi=T_R(\chi_R \psi)$, and therefore (using the relative form
boundedness of $V$ with respect to the Laplacian \cite[Theorem
X.19]{RS2}) we get, using Lemma~\ref{lem:SobTrace}, 
\begin{align*}
  \widetilde\rho_1(R) &= \| T_R \psi \|_{L^2(\mathbb{S}^2
  \times{\mathbb R}^{3N-3})}^2 
  \leq R^{-2} \langle \chi_R \psi, (-\Delta+1) \chi_R
  \psi\rangle \\
  &\leq c R^{-2}\langle\chi_R\psi,(H+1)\chi_R\psi\rangle\\ 
  &=cR^{-2}\langle\psi,\big(\tfrac{1}{2}(\chi_R^2 H
  +H\chi_R^2)+\chi_R^2+|\nabla \chi_R|^2 \big)\psi\rangle \\ 
  &=cR^{-2}\langle\psi,\big((E+1)
  \chi_R^2+|\nabla \chi_R|^2 \big)\psi\rangle\\
  &\leq c' R^{-2} \int_{\{{\mathbf x}\,:\, |x_1| > R-1\}}
  |\psi({\mathbf x})|^2\,d{\mathbf x}\,.
\end{align*}
{F}rom this \eqref{eq:IndivTerms} follows, and therefore
Lemma~\ref{lem:A} is proved. 
\end{proof}

Next, we define an operator \(\mathcal E_R\), which will
harmonically extend functions defined on
\(\mathbb{S}^2\times\R^{3N-3}\) (see \eqref{eq:harmextISharm} and
\eqref{eq:ResTextIsId} below).

We define by \(Y_{\ell,m}(\omega)\) the normalised (in \(L^2(\mathbb
S^2)\)) real valued spherical harmonics of degree \(\ell\),
\(\ell\in{\mathbb N}_0\), with \(m=0,1,\ldots,2\ell+1\). Then
\(\{Y_{\ell,m}\}_{\ell,m}\) constitutes an orthonormal basis in
\(L^2(\mathbb S^2)\); they are the eigenfunctions for \(\mathcal
L^2\), the Laplace-Beltrami operator on \(\mathbb S^2\) (\({\mathcal
  L}^2Y_{\ell,m}=\ell(\ell+1)Y_{\ell,m}\)).

Since \(L^2(\mathbb{S}^2\times{\mathbb R}^{3N-3})\cong L^2(\mathbb
S^2; L^2({\mathbb R}^{3N-3}))\), we have, for \(\phi\in
L^2(\mathbb{S}^2\times{\mathbb R}^{3N-3})\), 
\begin{align*}
  \phi(\omega, \hat{\bf
    x}_1)&=\sum_{\ell=0}^{\infty}\sum_{m=0}^{2\ell+1} 
  Y_{\ell,m}(\omega)\phi_{\ell,m}(\hat{\bf x}_1)\,, \\
  \phi_{\ell,m}(\hat{\bf x}_1) &=
  \int_{{\mathbb S}^2}Y_{\ell,m}(\omega)\phi(\omega,\hat{\bf
    x}_1)\,d\omega\,.
\end{align*}
Note that,  with \(\mathcal F\) the Fourier-transform on \(L^2(\mathbb
R^{3N-3})\),
\begin{align}
  \phi_{\ell,m}(\hat{\bf x}_1) = \frac{1}{(\sqrt{2\pi})^{3N-3}}\int_{{\mathbb
  R}^{3N-3}} e^{i\hat{\bf k}_1\cdot \hat{\bf x}_1}
  (\mathcal F{\phi}_{\ell,m})(\hat{\bf k}_1)\,d\hat{\bf k}_1\,.
  \end{align}
Define
\begin{align}
\label{def:E_R}
  [{\mathcal E}_R &\phi](r, \omega, \hat{\bf x}_1)\\&=
  \frac{1}{(\sqrt{2\pi})^{3N-3}}\sum_{\ell=0}^{\infty}\sum_{m=0}^{2\ell+1}
  \int_{{\mathbb R}^{3N-3}} 
  \!\!\!\!
  Y_{\ell,m}(\omega) e^{i\hat{\bf k}_1\cdot \hat{\bf x}_1} (\mathcal
  F\phi_{\ell,m})(\hat{\bf k}_1) f_{\ell, \hat{\bf k}_1,R}(r)\,d\hat{\bf
    k}_1,
  \nonumber 
\end{align}
where $f = f_{\ell, \hat{\bf k}_1,R}$ is the solution (given by
Lemma~\ref{lem:maxprinciple} in Appendix~\ref{app:Whittaker}) to the
equation 
\begin{align}\label{eq:Whittaker}
  f''+ \frac{2}{r} f'- \big[\frac{\ell(\ell+1)}{r^2} 
  + \hat{\bf k}_1^2\big] f = 0\ , \quad r \in (R, +\infty),
\end{align}
satisfying $f(R)=1$, \(|f(r)|\le1\) for all \(r\ge R\).

Using again Lemma~\ref{lem:maxprinciple} we find that
$$
  {\mathcal E}_R \phi \in L^{\infty}\big([R, \infty);
  L^2(\mathbb{S}^2\times{\mathbb R}^{3N-3})\big).
$$
Furthermore, by the definition of $f_{\ell, \hat{\bf k}_1,R}$,
${\mathcal E}_R \phi$ satisfies (in the sense of distributions)
\begin{align}\label{eq:harmextISharm}
  \Delta {\mathcal E}_R \phi = 0 \quad \text{ in } \quad
  \big\{ {\mathbf x}\in\mathbb R^{3N}\,\big|\, |x_1| > R \big\}.
\end{align}
We also have that
\begin{align}\label{eq:ResTextIsId}
  T_R {\mathcal E}_R \phi = \phi.
\end{align}
By analogy with $\Omega(R_1,R_2)$ from \eqref{eq:Omega}, we define 
\begin{align}
  \widetilde{\Omega}(R_1,R_2) = \big\{{\mathbf x} \in {\mathbb
  R}^{3N} \,\big| \, R_1 < |x_1| < R_2 \big\}.
\end{align}
With this notation we have the following \(L^2\) - bound:
\begin{lemma}\label{lem:B} 
For all \(R>0\) and \(\phi\in L^2({\mathbb
S}^2\times\mathbb{R}^{3N-3})\),  
\begin{align}
  \|{\mathcal E}_R \phi \|_{L^2(\widetilde{\Omega}(R,3R))}
  \leq 3R^{3/2} \| \phi \|_{L^2(\mathbb{S}^2
  \times{\mathbb R}^{3N-3})}.
\end{align}
\end{lemma} 
\begin{proof}
Using the definition (see \eqref{def:E_R}) of ${\mathcal E}_R$, the
Plancherel theorem, and Lem\-ma~\ref{lem:maxprinciple}, we get
\begin{align*}
  &\int_R^{3R}\Big(\int_{\mathbb R^{3N-3}} \int_{\mathbb S^{2}}
  |{\mathcal E}_R 
  \phi(r,\omega,\hat{\bf x}_1)|^2\,d\omega\,d\hat{\bf x}_1\Big) r^2\,dr
  \\
  &=(2\pi)^{3-3N}\int_R^{3R} \Big(\sum_{\ell=0}^{\infty}\sum_{m=0}^{2\ell+2}
  \int_{{\mathbb R}^{3N-3}} 
  \int_{{\mathbb S}^{2}} \big|Y_{\ell,m}(\omega)\big|^2\,
  \big|(\mathcal F \phi_{\ell,m})(\hat{\bf k}_1)\big|^2
  \\&\qquad\qquad\qquad\qquad\qquad\qquad\qquad\qquad
  \times|f_{\ell,\hat{\bf k}_1,R}(r)|^2\,d\omega\,d\hat{\bf k}_1\Big) r^2\,dr \\ 
  &\leq\| \phi \|_{L^2(\mathbb{S}^2\times{\mathbb R}^{3N-3})}^2
  \int_R^{3R}  r^2\,dr. 
\end{align*}
This implies the conclusion of the lemma.
\end{proof}
\begin{lemma}\label{lem:C-Sym}
Let $\psi$ be an eigenfunction of $H_{N_1,N_2}$ with eigenvalue $E$
and assume that $E<\inf \sigma_{{\rm ess}}(H_{N_1,N_2})$. 

Then there exist constants
$R_0,c>0$ such that for all $R>R_0$ and all $j \in
\{1,\ldots, N\}$,
\begin{align}\label{eq:Casej-Sym}
  \int_R^{\infty} \widetilde{\rho}_j(r) r^2\,dr
  \leq
  c R^{3} \widetilde{\rho}_j(R).
\end{align}
In particular,
\begin{align}\label{eq:SumCase-Sym}
  \int_R^{\infty} \widetilde{\rho}(r) r^2\,dr
  \leq c R^{3} \widetilde{\rho}(R).
\end{align}
\end{lemma}
\begin{proof}
Clearly \eqref{eq:SumCase-Sym} follows from \eqref{eq:Casej-Sym} by
summation over $j$, so we will only prove
\eqref{eq:Casej-Sym}. Without loss of generality, we only consider
$j=1$ and therefore aim to prove
\begin{align}\label{eq:CaseN}
  \int_R^{\infty} \widetilde{\rho}_1(r) r^2\,dr
  \leq
  c R^{3} \widetilde{\rho}_1(R).
\end{align}

We define (with the previously defined operators ${\mathcal E}_R$, 
$T_R$, see \eqref{def:T_R} and \eqref{def:E_R})
$$
  u = {\mathcal E}_R T_R \psi,
$$
as a function on ${\mathbb R}^{3N}$. The function $u$ does not
necessarily satisfy the antisymmetry properties from
$\mathcal{Q}(H_{N_1,N_2})$. Therefore,  denote, for $x \in {\mathbb
R}^3$, by $u_x$ the function on ${\mathbb R}^{3N-3}$ defined by 
$$
  u_x(x_2,\ldots , x_{N}) = u(x,x_2, \ldots , x_{N}).
$$
We stress that $u_x$ is {\it not} a derivative of $u$. With this
definition $u_x$ has the useful symmetry property  
\begin{align}
  \label{eq:eq:ux}
  u_x \in \mathcal{Q}(H_{N_1-1,N_2})\,.
\end{align}
{F}rom Lemma~\ref{lem:B} we get the inequality
\begin{align}\label{eq:L2-rhoSym}
  \int_{\widetilde\Omega(R,3R)} |u({\mathbf x})|^2
  \,d{\mathbf x} \leq 9R^3 \|T_R \psi\|^2_{L^2(\mathbb{S}^2\times{\mathbb
  R}^{3N-3})}
  =9R^3 \widetilde{\rho}_1(R).
\end{align}
Define $H_{N_1,N_2}(R)$ as the operator obtained by restricting $H$ to
the space 
\begin{align*}
  {\mathcal H}&_{N_1,N_2}(R) :=\\ 
  &\Big( 
  W^{2,2}\big( {\mathbb R}^3 \setminus \overline{B(0,R)}\big) \cap 
  W^{1,2}_0\big( {\mathbb R}^3 \setminus \overline{B(0,R)}\big) \Big)
  \otimes 
  W^{2,2}_{N_1-1,N_2}({\mathbb R}^{3N-3})\,.
\end{align*}
That is, we impose Dirichlet conditions at radius $R$ on the first
electron coordinate, and the symmetry conditions on the last $N-1$
electron coordinates.  

Let $\varphi \in {\mathcal H}_{N_1,N_2}(R)$ be normalised (in
$L^2(\mathbb R^{3N})$). Then, since
$|x_1|\geq R$, it follows that
\begin{align*}
  \langle \varphi, H_{N_1,N_2}(R) \varphi \rangle
  &\geq
  \Big\langle \varphi, \Big\{\sum_{j=2}^{N} \big(-\Delta_{j} - 
  \frac{Z}{|x_j|}\big) 
  \\&\qquad\qquad\qquad
  +\sum_{2\leq j<k \leq N}
  \frac{1}{|x_j - x_k|}\Big\} \varphi \Big \rangle - \frac{Z}{R}.
\end{align*}
By the HVZ-theorem (see \cite[Theorem XIII.17']{RS4})
the term in $\langle\,\cdot\,,\, \cdot\, \rangle$ (on the right side)
is bounded below by  
$\inf \sigma_{{\rm ess}}(H_{N_1,N_2})$. Thus, for any $\epsilon >0$
there exists $R' > 0$ such that for all $R > R'$,
$$
  \inf\sigma(H_{N_1,N_2}(R)) > \inf \sigma_{{\rm ess}}(H_{N_1,N_2}) - \epsilon.
$$
Since, by assumption, $E < \inf \sigma_{{\rm ess}}(H_{N_1,N_2})$, the
operator $H_{N_1,N_2}(R)-E$ is invertible for $R$ sufficiently large,
i.e., for all $R \geq R_0$ for some \(R_0>0\).

Let $\zeta \in C^{\infty}({\mathbb R}), 0\leq\zeta\leq1$, be a 
function such that 
\begin{align}\label{eq:defZetaSym}
  \zeta(t) &= 1 \text{ for } |t| \leq 2,&
  \zeta(t) &= 0 \text{ for } |t| \geq 3.
\end{align}
With $\zeta$ as above and $R>0$, we define $\zeta_R(x_1, \ldots, x_N)
:=\zeta(|x_1|/R)$. 
Let $v:=\psi - \zeta_R u$. Then $T_R v =0$, so we see using \eqref{eq:eq:ux} that 
$v \in {\mathcal H}_{N_1,N_2}(R)$.
A calculation gives
\begin{align}
(-\Delta + V - E) v
=-(V-E)\zeta_R u + 2 \nabla \zeta_R \nabla u + (\Delta \zeta_R) u.
\end{align}
Since $H_{N_1,N_2}(R)-E$ is invertible and $v \in {\mathcal
  H}_{N_1,N_2}(R)$, we find 
$$
  v = (H_{N_1,N_2}(R)-E)^{-1}\big( -(V-E)\zeta_R u + 2 \nabla \zeta_R
  \nabla u + (\Delta \zeta_R) u \big).
$$
It is easy to see, using that \(V\) is
relatively bounded with respect to the Laplacian,
\eqref{eq:L2-rhoSym}, and the support properties of $\zeta_R$, 
that there exist $c,c'$ such that
\begin{align}\label{eq:L2vSym}
  \| v \|^2_{L^2(\{{\mathbf x}\,:\,R < |x_1|\})}
  \leq c\| u \|^2_{L^2(\{{\mathbf x}\,:\,R < |x_1|<3R\})}
  \leq c' R^3 \widetilde{\rho}_1(R).
\end{align}
Combining \eqref{eq:L2-rhoSym} and \eqref{eq:L2vSym} we get
\begin{align*}
  \| \psi \|^2_{L^2(\{{\mathbf x}\,:\,R < |x_1|\})} &=
  \| \zeta_R u + v \|^2_{L^2(\{{\mathbf x}\,:\,R < |x_1|\})} \\
  &\leq
  2\big( \| \zeta_R u \|^2_{L^2(\{{\mathbf x}\,:\,R <
  |x_1|\})} + \| v \|^2_{L^2(\{{\mathbf x}\,:\,R <
  |x_1|\})} \big) \\
  &\leq c R^3 \widetilde{\rho}_1(R).
\end{align*}
This is the inequality \eqref{eq:CaseN}. The proof of
Lemma~\ref{lem:C-Sym} is therefore finished.
\end{proof}

The estimate \eqref{eq:limsupSym} follows from \eqref{eq:sumofterms}
and \eqref{eq:uppersolid}. The lower bound \eqref{eq:liminfSym}
clearly follows from  Lemma~\ref{lem:C-Sym} upon inserting
\eqref{eq:lowersolid}. This finishes the proof of
Theorem~\ref{thm:lowerExpDensitySym}.
\appendix
\section{} \label{app:Whittaker}
\begin{lemma}\label{lem:maxprinciple}
For all $\ell\in {\mathbb N} \cup \{0\}$, all $R>0$ and all
\(\kappa \geq 0\), the equation   
\begin{align}\label{eq:WhittakerBIS}
  f''+ \frac{2}{r} f'-\big[\frac{\ell(\ell+1)}{r^2} + \kappa^2\big] f
  = 0, \quad f(R)=1 
\end{align}
has a solution $f$ vanishing at infinity and satisfying 
\begin{align*}
  |f(r)| \leq 1 \ \text{ for all } r \geq R.
\end{align*}
\end{lemma}
\begin{proof}
Actually, if $f$ is a solution of \eqref{eq:Whittaker}, then $rf$ is a
Whittaker function (see \cite{AbraSte} for details). This implies that
a solution $f$ exists vanishing at infinity. 

Define, for $x \in {\mathbb R}^3$, $u(x):= f(|x|)$, then $u$ satisfies 
\begin{align*}
  -\Delta u + W u &= 0, & u\big |_{|x|=R} &= 1
\end{align*}
with $W(x) = \frac{\ell(\ell+1)}{|x|^2} + \kappa^2\geq 0$. 
By Kato's inequality \cite[Theorem X.27]{RS2}  we get
\begin{align}
 -\Delta |u| + W |u| &\leq 0.
\end{align}
Let
$v_{\kappa}$, $\kappa\geq 0$, be the function on ${\mathbb R}^3\setminus \{0\}$,  
\begin{align*}
  v_{\kappa}(x) = \frac{R}{|x|} e^{-\kappa(|x|-R)}. 
\end{align*}
Then
\begin{align*}
  -\Delta v_{\kappa} + \kappa^2 v_{\kappa} & = 0, & v_{\kappa} \big
   |_{|x|=R} &= 1, & \text{ and } v_{\kappa}(x) \leq 1\, \text{ for } |x| \geq R.
\end{align*}
So
\begin{align*}
  (-\Delta + W)(v_{\kappa} - |u|) &\geq
  \frac{\ell(\ell+1)}{|x|^2}v_{\kappa} \geq 0, \\ 
  (v_{\kappa} - |u|)\big |_{|x|=R} &= 0.
\end{align*}
The maximum principle (see e.\ g. \cite[Theorem
8.1]{Gilbarg-Trudinger}) implies (since $f$ and $v_{\kappa}$ vanish at
infinity) that for all $|x| \geq R$,  
\begin{align}\label{eq:MaxPrincUpper}
  |u(x)| \leq v_{\kappa}(x). 
\end{align}
This implies the statement of Lemma~\ref{lem:maxprinciple}.
\end{proof}

\begin{acknowledgement}
Parts of this work have been carried out at various
institutions, whose hospitality is gratefully acknowledged:
Mathematisches Forschungs\-institut Ober\-wolfach (SF, T\O S),
Erwin Schr\"{o}\-dinger Institute (SF, T\O S), Universit\'{e}
Paris-Sud (T\O S), the IH\'ES (T\O S), and the International Newton
Institute (SF, MHO, THO). Financial support from the 
European Science Foundation Programme {\it Spectral Theory and Partial
Differential Equations} (SPECT), and EU IHP network {\it Postdoctoral
Training Program in Mathematical Analysis of Large Quantum Systems},
contract no.\ HPRN-CT-2002-00277, is gratefully acknowledged.
T\O S was partially supported by the embedding grant from The Danish
National Research Foundation: Network in Mathematical Physics and
Stochastics, and by the European Commission through its 6th Framework
Programme {\it Structuring the European Research Area} and the
contract Nr. RITA-CT-2004-505493 for the provision of Transnational
Access implemented as Specific Support Action.
\end{acknowledgement}

\providecommand{\bysame}{\leavevmode\hbox to3em{\hrulefill}\thinspace}
\providecommand{\MR}{\relax\ifhmode\unskip\space\fi MR }
\providecommand{\MRhref}[2]{%
  \href{http://www.ams.org/mathscinet-getitem?mr=#1}{#2}
}
\providecommand{\href}[2]{#2}

\end{document}